# Diquark Approach to Calculating the Mass and Stability of H - Dibaryon

A. R. Haghpayma[†]

*Department of Physics, Ferdowsi University of Mashhad*

*Mashhad, Iran*

## Abstract

Diquarks may play an important role in hadronic physics particularly near the phase transitions ( chiral , deconfinement points ), current lattice QCD determinations of baryon charge distributions do not support the concept of substantial u - d scalar diquark clustering as an appropriate description of the internal structure of nucleon. Thus vector diquarks are more favourable. By using of vector diquark ideas in the chiral limit diquark correlations in the relativistic region and imposing HF interactions between quarks in a vector diquark we calculated the mass of H dibaryon, also by use of tunneling method we simultaneously calculated its decay width.

## 1. INTRODUCTION

QCD is believed to be the underlying theory of the strong interaction which has three fundamental properties: asymptotic freedom, colour confinement, approximate chiral symmetry and its spontaneous breaking. in high energy level QCD has been tested up to 0.01 level.

The behaviour of QCD in the low energy is nonperturbative and the $SU_C(3)$ colour group structure is non-abelian. However, besides conventional mesons and baryons, QCD itself does not exclude the existence of the nonconventional states such as glueballs ( gg , ggg , ..... ) hybrid mesons ( $q\bar{q}$ g ), and other multi - quark states ( qqqq , qqqqq ).

Do other multiquark hadrons exist? 4q , 6q , 7q , ... , Nq; is there an upper limit for N ?, study of these issues will deepen our understanding of the low- energy sector of QCD.

It is very difficult to calculate the whole hadron spectrum from first principles in QCD, under such a circumstance, various models which are QCD - based and incorporate some important properties of QCD were proposed to explain the hadron spectrum and other low-energy properties for more details, see Ref [1] in which we have explained and discussed some features of them.

In 1977 a bound six-quark state (uuddss), the H -dibaryon, was predicted in a bag-mode lcalculation by Jaffe[2]. This state is the lowest SU(3) flavor singlet state with spin zero, strangeness -2 and $J^P = 0^+$

In the last twenty years many attempts to verify the existence and stability of this particle were under taken by means of various methods. Perturbative calculations included spin-dependent q-q interactions arising from one gluon exchang.(OGE)[2,3], instanton induced interactions(III)[and Goldstone boson exchange (GBE)[4].

The attractive q-q interaction is controlled via the potential $V_s \sim (\lambda_1^a \lambda_2^a)(s_1 s_2)$, where $\lambda_i^a$ denote SU(3)color(OGE)or flavor(III/GBE) generators. The resulting H-masses scattered in a range of a few hundred MeV around the 2231 MeV $\Lambda\Lambda$-threshold for strong decay.

One of the great open problems of intermediate energy physics is the question of existence or nonexistence of dibaryons. Early theoretical models based on SU(3) and SU(6) symmetries [5,6] and on Regge theory [4,5] suggest that dibaryons should exist. There is QCD-based models predict dibaryons with strangeness S 0, -1, and -2. The invariant masses range between 2 and 3 GeV [9,10,11,12,13,14,15,16,17,18,19]. The masses and widths of the expected 6-quark states differ considerably for these models. But it seems that all QCD inspired models predict dibaryons and none forbids them. Until now, about 30 years after the first predictions of the S -2 H-dibaryon by Jaffe [9] this question is still open.

The theory of quantum chromodynamics imposes no specific limitation on the number of quarks composing hadrons other than that they form color singlet states. Although only qqq and states have been observed, other combinations can form colour singlets. Jaffe has proposed that a six-quark state uuddss may have sufficient colour-magnetic binding to be stable against strong decay. Such a state, which Jaffe named H, would decay weakly, and the resultant long lifetime would allow the possibility of observing such particles in neutral beams.

Theoretical estimates of have varied widely, ranging from a deeply bound state with 2.10 GeV/ to a slightly unbound state with near the threshold, 2.23 GeV/.

All experimental efforts failed so far to identify dibaryons. A possible reason could be that the experimental resolutions and the statistical accuracies were not sufficient. In this context it should be mentioned that the search for S 0 and S -2 dibaryons has been most intense whereas S -1 dibaryons have recieved less attention, although the lowest lying S -1 dibaryon states are expected to be very narrow.

## 2. DIQUARK APPROACH

We consider H dibaryon which composed of uuddss, three vector ud - ud - ss diquarks, each of which is a colour antitriplet and is symmetric in flavour and spin and orbital space, this leads to a six - quark state which is colour singlet, we ignore the pauli principle for quarks in different diquarks in the limit that diquarks are pointlike, but two quarks in each diquark satisfy this principle.

Now by using of diquark ideas in the chiral limit diquark correlations in the relativistic region and imposing HF interactions between quarks in a diquark, we led to introducing a conventional Hamiltonian:

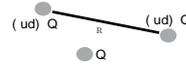

$$H(H) = T(H) + V^{CS}(H) \quad (1)$$

with

$$T(H) = \frac{\vec{\nabla}_R^2}{2m}$$

The orbital wave function is

$$\psi_m = N[a_R Y_{lm}(\hat{R})e^{-a^2 R^2/2}] \quad (2)$$

we calculated the masses of vector diquarks using colour - spin interactions as HF intraction between quarks in a diquark.

If we take over the results and consider a T= 450 Mev Kinetic energy as the binding energy for the H dibaryon the mass of it would be equal to the sum of the $\Xi^0$ and N masses.

Since the diquark masses ( e.x, vector or tensor, .... ) are smaller than the constituents, they are stable against decay near mass shell, in such a configuration, the diquarks are nearby and tunneling of one of the quarks between the two diquarks may take place.

We suppose that Decay widths of H is due to tunneling of one of the quarks between the two vector diquarks.

Thus in the decay process H → $\Xi^0$ N a (d) quark tunnels from a diquark ud to the other diquark ud to form a nucleon (udd) and an off-shell (u) quark which forms $\Xi^0$ with the other diquark .

The decay width of this process is

$$\Gamma_{\Theta^+} \simeq 5.0\, e^{-2S_0}\,\frac{g^2 g_A^2}{8\pi f_K^2}\,|\psi(0)|^2. \quad (3)$$

Which we have used WKB approximation for the tunneling amplitude and $\Delta E = (m_u + m_d - M_{ud})$ the $\psi(0)$ is the 1S wave function of quark - diquark at the origin and can be written as

$$\psi(0) = \frac{2}{a_0^{3/2}}\frac{1}{\sqrt{4\pi}}, \quad (4)$$

Where $a_0$ is the Bohr radius of the quark - diquark bound state and is $a_0 \simeq (2\bar{m}B)^{-1/2}$ where $\bar{m}$ = 250 Mev is the reduced mass and B is the binding energy of the quark - diquark bound state.

According to our model the Kinetic energy $T \simeq \frac{\vec{\nabla}_R^2}{2m} = \frac{3a^2}{4m} \simeq 450$ MeV this leads to $a = 444$ and then

$$r_0 = \langle R \rangle = \sqrt{\frac{5}{2a^2}} \simeq 0.003 \quad (5)$$

$g^2 = 3/03$, $g_A = 0.75$ from the quark model and $\Delta E = (m_u + m_d) - M_{ud}$ where $M_{ud} \simeq 520$ Mev in our model.

Inserting this values into Eq (3) we find

$$\Gamma_H \simeq 45 \text{ Mev} \quad (6)$$

[†]e-mail: haghpeima@wali.um.ac.ir

## 3. CONCLUSIONS

With the assumption that H dibaryon decays into $\Xi^0$ and N by tunneling of one (d) quark to (ud) diquark, we calculated its decay width $\Gamma_H \simeq 45$ Mev.

Thus the H dibaryon which is constructed by vector diquarks is unstable. There is other channels for H dibaryon decay for example into two $\Lambda$ baryon and one can estimate the decay width for them using this method.

Our theoretical results on the mass and width of H are in good agreement with many experimental results and one can use our vector diquark approach for calculatig the mass and width of other multiquark states.